\documentstyle[12pt,epsf]{article}

\begin{document}

\begin{center}

{\Large\bf Energy price risk management}

\vspace*{.6cm}

{\large Rafal Weron\footnotemark}

\footnotetext{E-mail: rweron@im.pwr.wroc.pl}

\vspace*{.3cm}

{\small
Hugo Steinhaus Center for Stochastic Methods,\\
Wroc{\l}aw University of Technology, 50-370 Wroc{\l}aw, Poland
}

\end{center}

\vspace*{.5cm}

\noindent
{\bf Abstract:} The price of electricity is far more volatile than that of other commodities 
normally noted for extreme volatility. Demand and supply are balanced on a knife-edge 
because electric power cannot be economically stored, end user demand is largely weather 
dependent, and the reliability of the grid is paramount. The possibility of extreme 
price movements increases the risk of trading in electricity markets. However, a number
of standard financial tools cannot be readily applied to pricing and hedging electricity 
derivatives. In this paper we present arguments why this is the case.\\


\vspace*{.5cm}

\section{Introduction}

Energy price risk management is still in its infancy compared to the more 
developed interest rate and foreign exchange markets \cite{fusaro,warrick,ww00}. 
However, we have to bear in mind that commodity markets are not anywhere near as
straightforward as financial markets. They have to deal with the added complexity 
of physical substance \cite{bll99}, which cannot simply be manufactured, transported 
and delivered, at the press of a button.

An innate energy industry conservatism coupled with highly profitable years caused stagnation 
despite the two oil price shocks of the 1970s. But the world oil price collapse of 1986
and the beginning of electric utility deregulation and privatization throughout the world 
are continuing to drive change in energy commodity markets \cite{fusaro,ICC,kurtz}.

In the wake of the recent price run-ups and defaults, managers have been forced to 
review their credit and counterparty risk policies. Traditional credit analysis has 
emphasized the financial risk associated with the failure of a buyer to pay for the 
goods purchased. Although this can be a concern in the power market, recent events 
have highlighted the substantial and perhaps less predictable market risk resulting 
from the failure to deliver by a seller \cite{calpx}. The defaults in late June 1998 
threw the US buyers into a superheated Midwest market, desperate for replacement power. 
This resulted in soaring prices that reportedly topped out at \$7,500 in real-time 
trading -- 300 times the average price of \$25/MWh!

The possibility of extreme price movements increases the risk of trading in electricity 
markets. Unfortunately a number of standard financial tools cannot be readily 
applied to pricing and hedging electricity derivatives. But before we explain why
let us briefly describe today's electricity markets.

\section{Electricity markets}

The deregulation of the electricity industry is giving way to a global trend toward 
the commoditization of electric energy \cite{green}. This trend has recently intensified 
in Europe and North America, where market forces have pushed legislators to begin removing 
artificial barriers that shielded electric utilities from competition. As a result,
during the last decade, we have witnessed a major explosion in the number of nontraditional 
power suppliers and financial engineers marketing electricity and electricity derivatives 
in the wholesale power markets. Only in the early four year period 1993-96 over 200 new 
marketers (qualified energy brokers) have appeared on the US electricity market \cite{ICC}.
Similarly, during the last six months more then 70 companies have obtained licenses to
trade electricity on the just liberalized Polish wholesale power market.

Organizations which have been used to long-term fixed price contracts are now 
becoming increasingly exposed to price volatility and, of necessity, are seeking
to hedge and speculatively trade to reduce their exposure to price risk.
The scenario in today's energy market is similar in many ways to the emergence of 
derivatives trading in the capital markets. From the modest beginnings 
in the late 1970s, financial markets have seen a massive explosion in the use of
derivative products. Starting with simple futures contracts and forward rate
agreements through swaps and on to increasingly ingenious and complex contracts.
The financial derivatives markets invented layer upon layer of new derivatives 
products using the basic building blocks to design tailor made hedges for 
customers \cite{coleman}.

Most derivatives markets begin with exchange traded futures. Global energy markets are 
no different. Heating oil futures appeared on the New York Mercantile Exchange (NYMEX) 
in November 1978 and futures on crude oil appeared there in March 1983. As in all other
markets options followed quickly -- on crude oil futures in 1986 and on heating oil a year
later. Commercial banks began providing commodity price risk management products in 1986,
when The Chase Manhattan Bank arranged the first oil swap \cite{ssw95}. 
Natural gas futures and OTC instruments began in 1990. And electricity futures contracts 
began trading on the world's most mature Scandinavian power market (Nord Pool -- 
the Nordic Power Exchange) in 1995 followed next year by the US (NYMEX) and 
Australian/New Zealand markets. 
First electricity options appeared on NYMEX in 1996. Last year Nord Pool introduced first 
exchange traded exotic options -- asian (average) options on electricity futures. 
Gas and electricity are now accelerating the change process of liquid trading and 
cross-energy commodity arbitrage. In effect, a conservative industry is continuing 
to be transformed through financial engineering.

\section{Arbitrage pricing}

Demand and supply of electricity are balanced on a knife-edge because 
electric power cannot be economically stored, end user demand is largely weather 
dependent (see Fig. 1), and the reliability of the grid is paramount. 
Relatively small changes in load or generation can cause large changes in price and all 
in a matter of hours, if not minutes. In this respect there is no other market like it. 

\begin{figure}[htbp]
\centerline{\epsfxsize=10cm \epsfbox{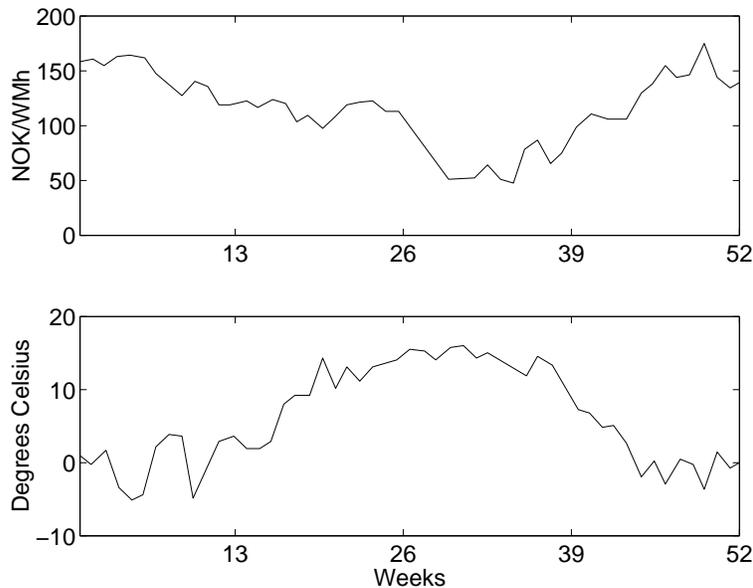}}
\caption{Nord Pool spot system price (top) and mean temperatures in Oslo, Norway (bottom) 
for the whole 1998 year. Clearly lower temperatures cause higher power consumption 
(heating) and thus raise electricity prices.
}
\end{figure}

Because of storage problems standard arbitrage type arguments cannot be used to price a number of
electricity derivatives \cite{ww00}. For example, when pricing a forward contract on crude oil, 
i.e. a contract for delivery of a specified amount of oil for a fixed price $K$ at a specified 
location and time in the future, we use the formula
$$
K=U (1+rT) + C,
$$
where $U$ is the current price of crude oil, $r$ is the risk-free 
interest rate, $T$ is the time to maturity of the contract, and $C$ is the so called 
{\it cost of carry} (a sum of storage, insurance, spoilage, and obsolescence costs). 
The formula can be derived by analyzing the following strategy \cite{ww00,ww98,hull97}:
\begin{itemize}
\item take a short position in the forward contract (i.e. agree to deliver oil for 
a fixed price $K$ at maturity) and take a loan from a bank to finance buying crude 
oil worth $U$ dollars,
\item store the oil until maturity (for time $T$) incurring the cost of carry $C$,
\item at maturity deliver the oil to the buyer of the forward contract for a fixed price $K$
and return the loan (with interest) to the bank.
\end{itemize}
The forward price $K$ should be such that from today's (the time we sign the contract) 
perspective no arbitrage is possible, i.e. we cannot make money without taking risk.
Thus the forward price should equal today's price ($U$) plus interest ($UrT$) 
paid to the bank for lending the money plus the cost of carry ($C$).
However, for contracts written on electricity the cost of carry is very large (or even
infinite) compared to the value of the delivered commodity and for this reason arbitrage 
type arguments cannot be used to price electricity derivatives. So what can we do? Well, 
we can either use other methods (eg. weather correlations, consumption prediction) for 
pricing such derivatives or use derivatives written not on electricity itself but on 
other derivatives.
The former method is used for pricing the first layer of derivatives -- forwards and
futures on electricity, whereas the latter for next layers -- options on futures 
(note that all exchange traded options are written on electricity futures and not 
electricity itself), swaptions, etc.

\section{Volatility}

In Figure 2 the average daily prices of the California Power Exchange 
(CalPX) spot market and their returns are plotted for the period April 1st, 1998 -- 
January 31st, 2000. Note that, unlike in the financial markets, electricity is traded 
every hour of the year -- including nights, weekends and holidays. Average daily price 
is a simple reference index constructed by adding up all 24 hourly prices during a day 
and dividing the sum by 24. One hour is the smallest time interval when prices can change, 
because in spot electricity trading prices are set constant for delivery of power during 
a certain hour. 

\begin{figure}[htbp]
\centerline{\epsfxsize=10cm \epsfbox{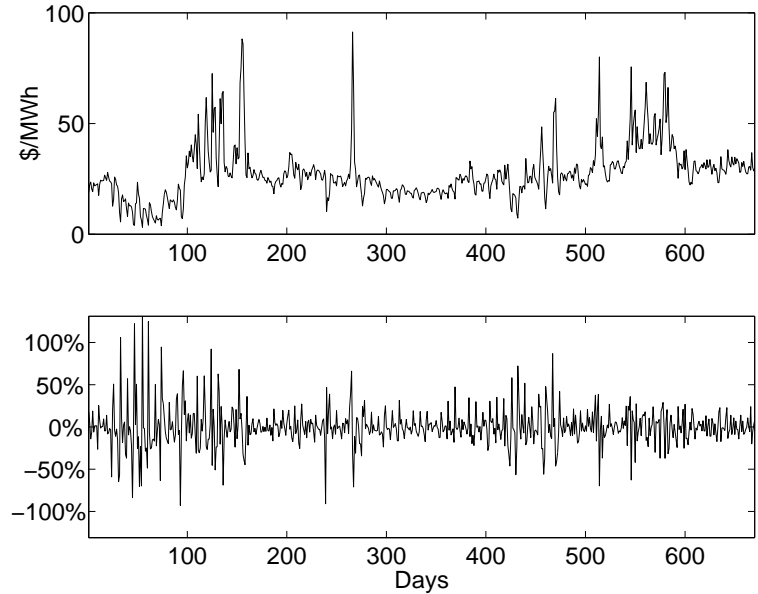}}
\caption{Average daily spot prices (top) and their returns (bottom) for the California
Power Exchange since the opening of the exchange (April 1st, 1998) till January 31st, 2000. 
}
\end{figure}

The price of electricity is far more volatile than that of other commodities normally 
noted for extreme volatility. Applying the classical notion of volatility -- the standard 
deviation of returns (i.e. logarithmic price changes: $r_t= \log x_{t+1} - \log x_t$), 
we obtain that measured on a daily scale for a series roughly one year in length: 
\begin{itemize}
\item treasury bills and notes have a volatility of less than 0.5\%,
\item stock indices have a moderate volatility of about 1-1.5\%,
\item commodities like crude oil or natural gas have higher volatilities (1.5-4\%),
\item very volatile stocks have volatilities not exceeding 4\%,
\item and electric energy has the highest volatility -- up to 30\%!
\end{itemize}
However, when measured on different time scales, electricity price volatility does not 
behave like that for financial instruments. For the data illustrated in Fig. 2
daily volatility is about 23\%, whereas monthly (30-day) volatility is about 33\%. 
This is much less then predicted by Brownian motion (the distance traveled 
by a particle is proportional to the square root of time) for which we would obtain
$23\% \times \sqrt{30} \approx 125\%$. Thus Black-Scholes type formulas \cite{ssw95,ww98,hull97} 
should in general overestimate premiums of long-term options written on electricity! 

Another feature of electricity price volatility is its seasonal character.
The daily and weekly seasonality of volatility can be illustrated \cite{o93,o97}
by the intra--weekly plot of mean absolute hourly price changes, see Fig. 3. 
The statistical week is divided into 168 hours from Monday 0:00-1:00 to Sunday 23:00-24:00. 
Each bar represents the mean absolute change in prices for every hour from the same 
week--days counted for energy prices from CalPX. The sampling period starts with the 
opening of the exchange (April 1st, 1998) and lasts until January 25th, 2000, so that 
we analyze 95 full weeks.

\begin{figure}[htbp]
\centerline{\epsfxsize=10cm \epsfbox{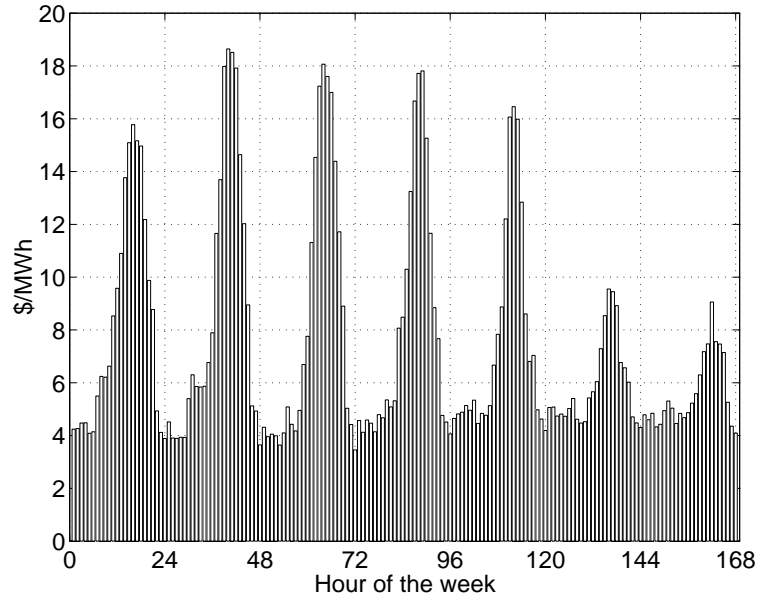}}
\caption{Intra--weekly plot of mean absolute hourly price changes for the CalPX spot
market. The statistical week is divided into 168 hours from Monday 0:00-1:00 to Sunday 
23:00-24:00.
}
\end{figure}

The patterns of volatility are clearly correlated to the on-peak/off-peak specification 
of the market. The lowest volatility is observed on the weekends and during night (off-peak) 
hours. However, unlike for the global interbank FX (currency) market \cite{o97}, 
the volatility during weekends is of the same order of magnitude as that for working days.
High volatility is observed during on-peak working day hours, with a maximum for hour 
15:00-16:00. Saturday has a similar volatility pattern, but on Sunday the maximum is 
postponed till 17:00-18:00.

\section{Autocorrelation of returns}

Seasonality of a time series of returns $r_t$ can be demonstrated by plotting 
the autocorrelation function \cite{beran94}
$$
{\bf acf}(r,k) = \frac{\sum_{t=k+1}^{N}
  (r_t-\bar{r})(r_{t-k}-\bar{r})}{\sum_{t=1}^{N} (r_t-\bar{r})^2},
$$
where $N$ is the sample length and
$$
\bar{r} = \frac{1}{N} \sum_{t=1}^{N} r_t,
$$
for different time lags $k$ as in Fig. 4. For electricity spot price returns there is 
a strong 7-day dependence which, when we think about it, is not that surprising. However, 
what is surprising is the fact that this dependence structure lasts almost forever 
(or as long as the analyzed data set)! For most financial data autocorrelation of returns
dies out (or more precisely: falls into the confidence interval of Gaussian random walk)
after 10-20 days and long-term autocorrelations are found only for squared returns 
or absolute value of returns \cite{ww98,o93,o97,bp97,ms99}.  

\begin{figure}[htbp]
\centerline{\epsfxsize=10cm \epsfbox{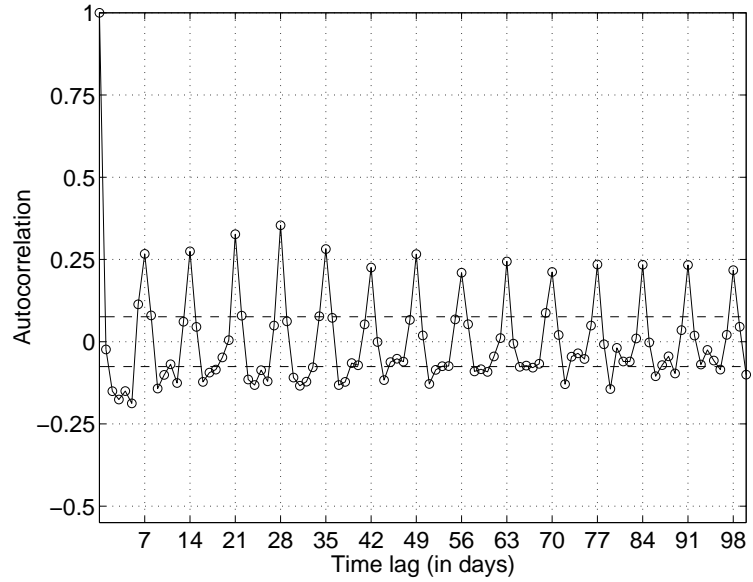}}
\caption{Lagged autocorrelation function for CalPX spot price returns (see Fig. 2).
Dashed horizontal lines represent the 95\% confidence interval of a Gaussian random walk.
}
\end{figure}

This 7-day cyclic correlation can be removed by differentiation, i.e. by constructing
the data series $z_t = r_{t+7} - r_t$. The lagged autocorrelation of $z_t$ is shown in 
Fig. 5. We can observe two evident outliers: for lag=1 day and for lag=7 days. Both
have negative correlations. This implies a strong mean-reverting property \cite{hull97} 
of the returns as was already suggested by the results of Section 3. 

\begin{figure}[htbp]
\centerline{\epsfxsize=10cm \epsfbox{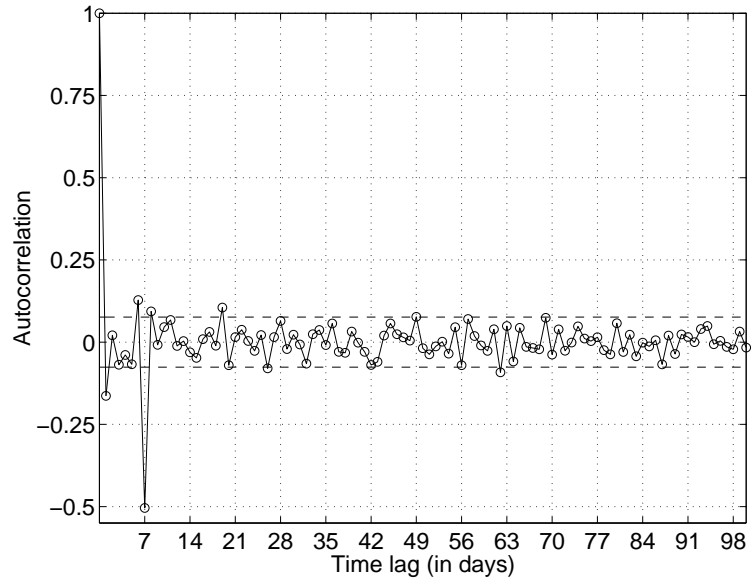}}
\caption{Lagged autocorrelation function for CalPX spot price returns after 
differentiation by 7 days. A strong mean-reverting property is revealed.
}
\end{figure}

\section{Final remarks}

As we have shown the price of electricity is far more volatile than that of other 
commodities normally noted for extreme volatility. However, the term structure of
volatility distinguishes electricity from most financial assets and forces us to use 
Black-Scholes type models with great care and a doze of skepticism. 
On the other hand, the mean-reverting property puts electricity in the same box as 
interest rates and suggests that the search for models of electricity price dynamics 
should be started with examining and calibrating certain interest rate models 
\cite{rb80,hw90}.

\end{document}